\documentclass[a4paper,12pt]{amsart}
\usepackage[T1]{fontenc}
\usepackage{mathabx}
\usepackage{dsfont}
\usepackage{color}
\usepackage{epsfig}
\usepackage{hyperref}
\usepackage{enumitem}
\usepackage{calc}
\usepackage{ifthen}
\pdfoutput=1
\usepackage{geometry}
\setlist[enumerate]{labelsep=*, leftmargin=1.5pc,
topsep=1ex plus0.5ex minus0.2ex,
itemsep=1ex plus0.5ex minus0.2ex,
font=\rmfamily,
font=\upshape}
\setlist[itemize]{labelsep=*, leftmargin=1.5pc,
topsep=1ex plus0.5ex minus0.2ex,
itemsep=1ex plus0.5ex minus0.2ex}
\newtheorem{Lem}{Lemma}[section]
\newtheorem{Pro}[Lem]{Proposition}
\newtheorem{Thm}[Lem]{Theorem}
\newtheorem{Cor}[Lem]{Corollary}
\theoremstyle{definition}

\newtheorem{Exa}[Lem]{Example}

\newtheorem*{Ack}{Acknowledgements}
\numberwithin{equation}{section}
\newcommand{\her}{^{\operatorname h}}
\newcommand{\tr}{{\rm tr}}
\newcommand{\id}{\mathds{1}}
\newcommand{\supp}{{\rm supp}}
\newcommand{\dd}{{\rm d}}
\newcommand{\cl}{{\rm cl}^{\rm rI}}
\newcommand{\ov}[1]{\overline{#1}}
\newcommand{\bC}{\mathds{C}}
\newcommand{\bN}{\mathds{N}}
\newcommand{\bR}{\mathds{R}}
\newcommand{\cA}{\mathcal A}

\newcommand{\cE}{\mathcal E}
\newcommand{\cF}{\mathcal F}
\newcommand{\cH}{\mathcal H}
\newcommand{\cM}{\mathcal M}

\newcommand{\cS}{\mathcal S}
\begin{document}
\bibliographystyle{plain}
\author{Stephan Weis, Andreas Knauf, Nihat Ay and Ming-Jing Zhao}
\title
{Maximizing the divergence from a hierarchical model of quantum states}
\maketitle
\thispagestyle{empty}
\pagestyle{myheadings}
\markleft{\hfill Divergence  from a hierarchical model\hfill}
\markright{\hfill S.\ Weis, A.\ Knauf, N.\ Ay, M.-J.\ Zhao\hfill}
\begin{abstract}
We study many-party correlations quantified in terms of the Umegaki
relative entropy (divergence) from a Gibbs family known as
a hierarchical model. We derive these quantities from the
maximum-entropy principle which was used earlier to define the
closely related irreducible correlation. We point out differences
between quantum states and probability vectors which exist 
in hierarchical models, in the divergence from a hierarchical
model and in local maximizers of this divergence. The differences are, 
respectively,
missing factorization, discontinuity and reduction of uncertainty. We 
discuss global maximizers of the mutual information of separable qubit
states.
\end{abstract}
\vspace{.3in}
\par
Index Terms:
many-party correlation,
maximum-entropy principle,
hierarchical model,
irreducible correlation,
mutual information,
multi-information,
factorization,
discontinuity,
maximizer,
separable state
\vspace{.2in}
\par
AMS Subject Classification:  62H20, 62F30, 94A17, 81P16, 81P45
\vspace{.3in}
%
%
%
%
\section{Introduction}
\label{sec:intro}
\par
In this article we quantify many-party correlations in the state of a 
composite quantum system which can not be observed in subsystems composed 
of less than a given number of parties. One of us \cite{Ay-Preprint} 
has quantified stochastic interactions in terms of a distance from 
non-interacting states. Following this idea, we replace in the present 
context the non-interacting states by states which are fully 
described by their restriction to selected subsystems. For a definition
of the latter states the maximum-entropy principle
was suggested earlier \cite{Amari,Linden-etal2002,Zhou2008} because it 
solves the inverse problem to reconstruct a global state from 
subsystem states and it offers also a natural scale of 
many-party correlation in terms of the gap to the maximal entropy 
value.
Mathematical deduction leads from here to the conception 
\cite{Ay-Preprint,Amari,AyOlbrichBertschingerJost,Zhou2009,Weis-MaxEnt14}
that many-party correlation should be quantified in terms of the 
divergence (which is an asymmetric distance) from a family of Gibbs 
states which we will call hierarchical model in the sense of 
\cite{Lauritzen}.
\par
We are considering a composite system of $N\in\bN$ units, parties, 
particles, etc.\
$[N]:=\{1,\ldots,N\}$. Tacitly, probability vectors on a finite space
(classical case) are included in this discussion of quantum systems
because vectors can be embedded as diagonal matrices into a matrix
algebra (quantum case). We consider the algebra $\cM_d$ of complex
$d\times d$ matrices with identity $\id_d$, $d\in\bN$, and we endow it
with the Hilbert-Schmidt inner product $\langle a,b\rangle:=\tr(ab^*)$,
$a,b\in\cM_d$. Each unit $i\in[N]$ has a unit size $n_i\in\bN$ and a
C*-subalgebra $\cA_i\subset \cM_{n_i}$ such that $\id_{n_i}\in\cA_i$.
The composite system is described by the tensor product algebra
$\cA_{[N]}:=\cA_1\otimes\cdots\otimes\cA_N$.
\par
The simplest notion of correlation is the total correlation.
The corresponding set of states without any correlations
is the space of tensor product states
\begin{equation}\label{eq:F1}
\cF_1:=\{\rho_1\otimes\cdots\otimes\rho_N\mid
\mbox{ $\rho_i$ is a quantum state of unit $i\in[N]$}\}.
\end{equation}
Here a {\em state} of a quantum system with C*-algebra $\cA\subset \cM_d$,
$d\in\bN$, denotes a {\em density matrix} which is a positive
semi-definite matrix in $\cA$ of unit trace \cite{Ruelle}.
We observe the following.
\begin{itemize}
\item
The states in $\cF_1$ are totally uncorrelated in the sense that the
probability distribution of the measurement outcomes (with respect to
a projective \cite{Nielsen} or simple \cite{Amari-Nagaoka}
measurement) of an observable $a_1\otimes\cdots\otimes a_N$ has the
product form.
\item
Any distance of a quantum state from $\cF_1$ quantifies
correlations in the Aristotelian sense that the whole is more than
the sum of its parts, cf.\ \cite{Ay-Preprint}. Here a distance should 
be zero for points in $\cF_1$ 
and strictly positive otherwise.
\end{itemize}
\par
It is interesting to differentiate correlations between the number of
particles which interact.
An algebraic generalization from no correlation (\ref{eq:F1}) to $k$-party
interaction, $k\in\bN$, is unknown in the quantum setting, although it
exists classically as we recall in Sec.~\ref{sec:classical-factorization}.
The way out is the maximum-entropy principle \cite{Jaynes} which
also delivers a natural scale for correlations:
In Sec.~\ref{sec:maxent-to-div-intro} we define a quantity $c_k(\rho)$
capturing all correlations in a state $\rho$ in $\cA_{[N]}$ which can not
be observed in any $k$-party subsystem. Later in
Sec.~\ref{sec:hierarchical-models} we introduce the notion of hierarchical
model which allows to define interaction patterns of subsystems which are
more general than the class of $k$-party subsystems.
\par
Based on our earlier work 
\cite{Weis-support,Weis-topology,Weis-MaxEnt14,Weis-Knauf}
we recall in Sec.~\ref{sec:fundamentals} that the many-party correlation 
$c_k$ is just the divergence
\begin{equation}\label{eq:c=d}
c_k(\rho)=\inf\{D(\rho,\sigma)\mid\sigma\in\cE_k\},
\quad\mbox{$\rho$ a state in $\cA_{[N]}$}
\end{equation}
from the {\em Gibbs family}
\begin{equation}\label{eq:Ek}
\cE_k:=\{e^H/\tr(e^H)\mid H\in\cH_k\}
\end{equation}
of the $k$-local Hamiltonians $\cH_k$.
Here a {\em $k$-local Hamiltonian} \cite{Kempe,Cubitt}
is defined as a sum of tensor product terms $a_1\otimes\cdots\otimes a_N$
with at most $k$ non-scalar factors $a_i\in\cA_i\her$, $i\in[N]$, where
$\cA\her$ denotes the real space of self-adjoint matrices in a C*-algebra
$\cA\subset \cM_d$, $d\in\bN$. The Umegaki relative entropy which we call
{\em divergence} is an asymmetric distance between states $\rho,\sigma$
in $\cM_d$ defined by
\[
D(\rho,\sigma):=\tr\,\rho(\log(\rho)-\log(\sigma))
\]
if the kernel of $\sigma$ is included in the kernel of $\rho$,
otherwise $D(\rho,\sigma):=\infty$. The distance-like property of
$D(\rho,\sigma)\geq 0$ with equality if and only if $\rho=\sigma$
is well-known \cite{Wehrl1978,Nielsen}. 
\par
Related concepts in the literature include the notion of 
{\em $k$-body potential} in 
statistical mechanics \cite{Ruelle} which is similar to the notion of 
$k$-local Hamiltonian. The proof of (\ref{eq:c=d}) that the correlation 
$c_k$ equals the 
divergence from $\cE_k$ has been given in probability theory in
\cite{Amari,AyOlbrichBertschingerJost}. The quantum mechanical proof in
\cite{Zhou2009} works only for states of maximal rank while the proof in
\cite{Weis-MaxEnt14} is valid without rank restriction. 
\par
Some new results are pointed out in Secs.~\ref{sec:zero-temp-intro} 
and~\ref{sec:max-div-intro}. We remark in Sec.~\ref{sec:zero-temp-intro} 
that the step from maximal rank to non-maximal rank has a physical interpretation 
as a zero-temperature limit. This step entails phenomena like a missing 
factorization of maximum-entropy probability distributions and a 
discontinuity of quantum correlations. We do not know how reliable the 
algorithms \cite{Niekamp-Galla} are at discontinuities of the divergence 
from $\cE_k$. In Sec.~\ref{sec:max-div-intro} we address maximizers of 
correlation and we point out a curious reduction of uncertainty in quantum 
maximizers.
\par
The Gibbs family $\cE_1$ is known as the {\em independence model} and
the divergence of a state $\rho$ in $\cA_{[N]}$ from $\cE_1$ quantifies 
the total correlation. We show in Sec.~\ref{sec:multi-information} that
the divergence from $\cE_1$ can be written in the form
\begin{equation}\label{eq:c1=H}
c_1(\rho)=H(\rho_{\{1\}})+\cdots+H(\rho_{\{N\}})-H(\rho)
\end{equation}
where the $\rho_{\{i\}}$ are one-party marginals
(Sec.~\ref{sec:maxent-to-div-intro}) and
\[
H(\sigma):=-\tr\sigma\log(\sigma)
\]
denotes the {\em von Neumann entropy} of a state
$\sigma$ in $\cM_d$, $d\in\bN$. 
The right-hand side of (\ref{eq:c1=H}) is also known as
{\em multi-information} \cite{Ay-Knauf} and quantifies
the number of random bits needed to erase all correlations
between the units of a composite system \cite{Groisman}
if the base of the logarithm is two.
\par
Finally, 
we remark that the divergence from an exponential family plays a major role
in the context of the maximum likelihood estimation \cite{Csiszar-Matus2008}.
The relative entropy of entanglement \cite{VedralPlenioRippinKnight} is
analogously defined in terms of the divergence from the convex set of
non-entangled states. However, this set does not form an exponential family.
Therefore this entanglement measure can not be motivated in terms of the
maximum entropy principle, in contrast to the divergence representation
(\ref{eq:c=d}) of the correlation quantity $c_k$. From the
information-geometric perspective, it is more natural to apply the relative
entropy projection onto a convex set with respect to the first argument of
$D$, which is consistent with the work \cite{Bjelakovic-etal} on hypothesis
testing.
%
%
\subsection{Interaction patterns}
\label{sec:maxent-to-div-intro}
\par
The maxi\-mum-en\-tro\-py prin\-ci\-ple, in its statistical inference view
\cite{Jaynes}, is suitable to introduce particle numbers into quantum
many-party correlations. If information about a state is available in the
form of a constraint (imagine a subset containing the state) then the
state which maximizes the von Neumann entropy $H$ under the constraint is
considered \cite{Jaynes}
the least informative state representing the given information.
Our constraints will be quantum marginals. Denoting the algebra of the
subsystem of units in $\nu\subset[N]$ by the tensor product
$\cA_\nu:=\bigotimes_{i\in\nu}\cA_i$ with identity $\id_\nu$, the
{\em $\nu$-marginal} $\rho_\nu$ of a state $\rho$ in $\cA_{[N]}$ is
defined by the equations
\[
\langle \rho_\nu,a\rangle=\langle\rho,a\otimes\id_{[N]\setminus\nu}\rangle,
\quad
a\in\cA_\nu.
\]
If for some $k\in\bN$ the information consists of the marginals of
all $k$-party subsystems, that is subsystems composed of $k$ units,
of some global state $\rho$ in $\cA_{[N]}$ then we notice
\begin{itemize}
\item any two states compatible with the constraint are indistinguishable
on any subsystem composed of $k$ or less units;
\item a state in $\cA_{[N]}$ which is compatible with the
constraint and has less entropy than the maximal entropy $H_{\max}$
has additional information.
\end{itemize}
Since $\rho$ is compatible with the constraint, it is natural to quantify
the additional information in $\rho$ by
$c_k(\rho):=H_{\max}-H(\rho)$. We take
this information as a definition of {\em many-party correlations}: The
quantity $c_k(\rho)$ captures all correlations in $\rho$ which can not
be observed in any $k$-party subsystem.
\par
We remark that the very closely related quantity of
{\em irreducible $k$-party correlation}
\cite{Linden-etal2002,Zhou2008} is defined by
$C_k(\rho):=c_{k-1}(\rho)-c_k(\rho)$ and  quantifies all correlations
which can be observed in the $k$-party subsystems but not in the
$(k-1)$-party subsystems. For example the irreducible three-party
correlation $C_3$ can be used to distinguish the genuine 3-party
correlation from 2-party correlation, like three-tangle in
\cite{Coffman-etal2000}. But entanglement is just one kind of quantum
correlation, so the quantity $C_3$ is different
from three-tangle. For the case of probability distributions
see for example
\cite{Kahle-Olbrich,AyOlbrichBertschingerJost}.
%
%
\subsection{Non-maximal rank phenomena}
\label{sec:zero-temp-intro}
\par
The step from maximal rank to non-maxi\-mal rank is crucial in ultra-cold
physics, for example in condensed matter physics \cite{Sachdev2014,Wen2004}
or adiabatic quantum computation \cite{Pachos2012}, because non-maximal
rank states are zero-temperature limits of Gibbs states in the sense
of $e^{-\beta H}/\tr(e^{-\beta H})$ for $\beta\to\infty$.
Mathematical phenomena of non-maximal rank appear in 
Sec.~\ref{sec:classical-factorization} in the context of higher factorization 
$\cF_1\subset\cdots\subset\cF_N$ by generalizing (\ref{eq:F1}). Higher 
factorization is unknown in the quantum case
but consequences may generalize from classical to quantum
systems, who knows? We anticipate that the inclusions
$\cE_k\subset\cF_k\subset\overline{\cE_k}$ are strict ($\overline{\cE_k}$
denotes norm closure) for $k\geq 2$. In a three-qubit quantum system it is
known that the divergence from $\cE_2$ is discontinuous
at the GHZ state \cite{Weis-MaxEnt14,RSSW}. This is indeed a very
pronounced irregularity and related phenomena have been suggested as 
signatures of quantum phase transitions \cite{CJLPSYZZ}. In the classical 
case the divergence from $\cE_k$ is continuous for all $k\in\bN$
\cite{Weis-topology}. We will return to the continuity problem
in Sec.~\ref{sec:fundamentals}.
%
%
%
%
\subsection{Maximizing the divergence}
\label{sec:max-div-intro}
\par
We have studied maximizers of the divergence from Gibbs families 
in the classical case for example in \cite{Ay,Ay-Knauf}. The latest 
result in the area is \cite{Rauh}. Two of us 
\cite{Weis-Knauf,Weis-topology}
have shown that quantum maximizers have properties analogous to the 
following classical ones provided in \cite{Ay}:
\begin{itemize}
\item
A local maximizer of the divergence from a Gibbs family $\cE$ is the
conditional distribution of its projection to $\cE$;
\item
a local maximizer of the divergence from $\cE$ is supported on
a set of size of at most $\dim_\bR(\cE)+1$.
\end{itemize}
We prove in Sec.~\ref{sec:local-maximizers} that the upper bound on
the support size improves in the quantum setting to $\sqrt{\dim_\bR(\cE)+1}$ 
because the state space of an $n$-level quantum system has
dimension $n^2-1$ compared to $n-1$ which is the dimension of the probability
simplex. For example, if all $N\in\bN$ units of a composite system have the
same unit size $n\in\bN$, then the independence model $\cE_1$ has
dimension $N(n-1)$ in the classical case and $N(n^2-1)$ in the quantum case
of a full matrix algebra.
Therefore, a local maximizer of the multi-information has support at
most ${\mathcal O}(N)$ respectively ${\mathcal O}(\sqrt{N})$, see the 
paragraph of (\ref{eq:equal-units-cl}). In a loose analogy,
if the classical bound was sharp, these bounds confirm that quantum systems
are less uncertain than classical systems
\cite{BenattiHudetzKnauf,CafaroGiffinLupoMancini}. In both cases 
we have an exponential reduction from the complete randomness with corresponding 
support size $n^N$.
\par
Global maximizers are less coherent in the classical-quantum
comparison. The classification of global maximizers of the
multi-in\-for\-ma\-tion \cite{Ay-Knauf} in the classical setting is
not valid in the quantum setting due to the entanglement. However,
we demonstrate in Sec.~\ref{sec:global-maximizers} that the methods
in \cite{Ay-Knauf} are helpful to understand maximizers of the
mutual information of separable qubit states.
%
%
%
%
\section{Factorization of probability distributions}
\label{sec:classical-factorization}
\par
We recall from \cite{Geiger,Develin} that the set of probability vectors
with at most $k$-party interactions has several algebraic representations.
Loopholes in the representations are explained by examples from \cite{Kahle}
and by proving their minimality.
\par
Let us associate to each unit $i\in[N]$ a state space $X_i$
which is an arbitrary set of cardinality equal to the unit size $n_i$
defined earlier. The composite system has the
state space $X_1\times\cdots\times X_N$. For any subset $\nu\subset[N]$
we consider a subsystem $X_{\nu}:=\bigtimes_{i\in\nu}X_i$ and for
any tuple $x=(x_1,\ldots,x_n)\in X_{[N]}$ its restriction
$x_\nu:=(x_i)_{i\in\nu}$ to the subsystem. We denote the probability simplex
over a finite set $X$ by
\[\textstyle
\Delta(X):=\{p\in\bR^X\mid \forall i\in X:p(i)\geq 0, \sum_{i\in X}p(i)=1\}.
\]
When switching to the notation of quantum systems in Sec.~\ref{sec:intro}
we tacitly identify $\bC^{X\nu}\cong\cA_\nu$ for subsets of units
$\nu\subset[N]$. Then $\Delta(X_\nu)$ is the set of states in $\cA_\nu$.
\par
A probability vector $p\in\Delta(X_{[N]})$ {\em factorizes} with respect to
$k$-party subsystems, $k\in\bN$, if there are functions
$\psi_\nu\in\bR^{X_\nu}$, $\nu\subset[N]$, $|\nu|=k$, such that
\begin{equation}\label{eq:factor-part-no}\textstyle
p(x)=\prod_{\nu\subset[N],|\nu|=k}\psi_\nu(x_\nu),\quad
x\in X_{[N]}.
\end{equation}
Let us denote by $\cF_k$ the set of all probability vectors with
(\ref{eq:factor-part-no}). Notice that the definition of $\cF_1$ is
consistent with (\ref{eq:F1}) in the classical case.
\par
We follow \cite{Geiger} by working out Lemma~\ref{lem:feasible-cond}.
Thereby we meet two representations of $\cF_k$. The lemma is a condition
for the inclusion of a probability vector into $\cF_k$ in terms of
the support. Using the set of $k$-party subsystem states
$I_k:=\bigcup_{\nu\subset[N],|\nu|=k}\{(\nu,x)\mid x\in X_\nu\}$
we define a matrix with rows indexed by $I_k$ and columns indexed by
$X_{[N]}$
\begin{equation}\label{eq:A-matrix-of-U}
a_{(\nu,y),x}:=\left\{
\begin{array}{rl}1 & \text{if }x_\nu=y\\0 & \text{else}\end{array}
\right.,
\quad (\nu,y)\in I_k, x\in X_{[N]}.
\end{equation}
See Example~\ref{ex:higher-interaction} for three bits and $k=2$.
Notice for all $x\in X_{[N]}$ that $\sum_{i\in I_k}a_{i,x}={N\choose k}$
holds. The matrix (\ref{eq:A-matrix-of-U}) defines a monomial map
\[
\Phi:[0,\infty)^{I_k}\to[0,\infty)^{X_{[N]}},
\quad
t\mapsto(\textstyle\prod_{i\in I_k} t(i)^{a_{i,x}})_{x\in X_{[N]}}
\]
where we agree on $0^0=1$ and $0^\alpha=0$ for $\alpha>0$. It is easy
to prove for $p\in\Delta(X_{[N]})$ that $p$ lies in $\cF_k$ if and only
if $p$ belongs to the image of $\Phi$. To get a second representation
of $\cF_k$ we define a family of functions
$r_\theta(x):=\exp(\sum_{i\in I_k}\theta(i)a_{i,x})$, $x\in X_{[N]}$,
with family parameter $\theta\in[-\infty,\infty)^{I_k}$. If
$\theta\in[-\infty,\infty)^{I_k}$ satisfies the condition
\begin{equation}\label{eq:normalizable}\textstyle
r_\theta(x)>0 \mbox{ holds for at least one }x\in X_{[N]}
\end{equation}
then a probability vector $p_\theta:=Z(\theta)^{-1}r_\theta$ is
defined where $Z(\theta)$ is for normalization. It is easily proved
that the set of constructed probability vectors $p_\theta$ is the
intersection of $\Delta(X_{[N]})$ with the image of $\Phi$.
\par
The {\em support} of a vector $v\in \bR^X$ indexed by a finite set $X$ 
is defined by $\supp(v):=\{x\in X\mid v(x)\neq 0\}$. The column of the matrix
(\ref{eq:A-matrix-of-U}) with column label $x\in X_{[N]}$ will be
written $a_x:=(a_{i,x})_{i\in I_k}$. We call a
non-empty subset $F\subset X_{[N]}$ {\it $k$-feasible} \cite{Geiger} if
\[\textstyle
\supp(a_x)\not\subset\bigcup_{y\in F}\supp(a_y)
\quad\mbox{holds for all } x\in X_{[N]}\setminus F.
\]
It is easy to see that a non-empty subset $F\subset X_{[N]}$ is
$k$-feasible if and only if $F$ is the support set of a vector
$r_\theta(x)$ for some $\theta\in[-\infty,\infty)^{I_k}$ satisfying
(\ref{eq:normalizable}). Restriction to $\theta\in\{-\infty,0\}^{I_k}$
gives the following.
\begin{Lem}\label{lem:feasible-cond}
The uniform probability vector supported on a non-empty subset
$F\subset X_{[N]}$ belongs to $\cF_k$ if and only if $F$ is
$k$-feasible.
\end{Lem}
\par
Notice that (\ref{eq:normalizable}) implies inclusions between
$\cF_k$ and the Gibbs family $\cE_k$ of the $k$-local Hamiltonians
(\ref{eq:Ek}):
\begin{equation}\label{eq:fact-in-closure}
\cE_k\subset\cF_k\subset\overline{\cE_k}.
\end{equation}
We recall a representation of $\overline{\cE_k}$
in Thm.~3.2 in \cite{Geiger} (unknown in the quantum case) where
$\overline{\cE_k}$ is the intersection of the probability
simplex $\Delta(X_{[N]})$ and of a {\it non-negative toric variety}
defined as the set of all vectors $s\in[0,\infty)^{X_{[N]}}$ such
that we have
\[\textstyle
\prod_{x\in X_{[N]}}s(x)^{u(x)}
=\prod_{x\in X_{[N]}}s(x)^{v(x)}
\]
for all $u,v\in \bN_0^{X_{[N]}}$ where $u-v$ lies in the kernel of
the matrix (\ref{eq:A-matrix-of-U}).
\par
Let us give an example to see why $\cF_k$ is not closed
for $k\geq 2$ and let us prove minimality of the example.
\begin{Exa}
\label{ex:higher-interaction}
Let $k,N\in\bN$ and $N>k\geq 2$. Then $\overline{\cE_k}\setminus\cF_k$
is non-empty. For simplicity we consider $N=k+1$ bits. The subset
\[
Y:=\{(x_1,\ldots,x_N)\mid x_i=0 \text{ for all but one } i\in[N]\}
\]
of $X_{[N]}=\{0,1\}^N$ is not feasible. So
Lemma~\ref{lem:feasible-cond} proves that the uniform probability vector
supported on $Y$ does not lie in $\cF_k$. On the other hand, the support sets
of distributions in $\overline{\cE_k}$ include all subsets of size $2^k-1$ by
Theorem~14 in \cite{Kahle}. Since $2^k-1\geq N$ holds for $k\geq 2$ and since
$Y$ has $N$ elements, the uniform probability
vector supported on $Y$ lies in $\overline{\cE_k}$. For $N=3$ the matrix
(\ref{eq:A-matrix-of-U}) is\\[\baselineskip]
\centerline{\scalebox{0.71}{
$\begin{array}{r|cccccccc}
 & (0,0,0) & (0,0,1) & (0,1,0) & (0,1,1) &
 (1,0,0) & (1,0,1) & (1,1,0) & (1,1,1)\\\hline
\{1,2\},(0,0) & 1 & 1 & 0 & 0 & 0 & 0 & 0 & 0 \\
\{1,2\},(0,1) & 0 & 0 & 1 & 1 & 0 & 0 & 0 & 0 \\
\{1,2\},(1,0) & 0 & 0 & 0 & 0 & 1 & 1 & 0 & 0 \\
\{1,2\},(1,1) & 0 & 0 & 0 & 0 & 0 & 0 & 1 & 1 \\
\{2,3\},(0,0) & 1 & 0 & 0 & 0 & 1 & 0 & 0 & 0 \\
\{2,3\},(0,1) & 0 & 1 & 0 & 0 & 0 & 1 & 0 & 0 \\
\{2,3\},(1,0) & 0 & 0 & 1 & 0 & 0 & 0 & 1 & 0 \\
\{2,3\},(1,1) & 0 & 0 & 0 & 1 & 0 & 0 & 0 & 1 \\
\{1,3\},(0,0) & 1 & 0 & 1 & 0 & 0 & 0 & 0 & 0 \\
\{1,3\},(0,1) & 0 & 1 & 0 & 1 & 0 & 0 & 0 & 0 \\
\{1,3\},(1,0) & 0 & 0 & 0 & 0 & 1 & 0 & 1 & 0 \\
\{1,3\},(1,1) & 0 & 0 & 0 & 0 & 0 & 1 & 0 & 1
\end{array}$}.}\\[\baselineskip]
The equation of the non-negative toric variety which represents
$\overline{\cE_2}$ is known \cite{Develin} and equals
$\begin{smallmatrix}
p(0,0,0)p(0,1,1)p(1,0,1)p{(1,1,0)}
=p{(0,0,1)}p{(0,1,0)}p{(1,0,0)}p{(1,1,1)}
\end{smallmatrix}$.
\end{Exa}
\par
The cardinality of the non-feasible set $Y$ in
Example~\ref{ex:higher-interaction} is minimal.
\begin{Lem}
Let $l,k,N\in\bN$ and $1\leq l\leq k\leq N$. Then every subset of $X_{[N]}$
of cardinality $l$ is $k$-feasible.
\end{Lem}
{\it Proof:}
For any $x\in X_{[N]}$ we denote the support of the $x$-th column of the
matrix (\ref{eq:A-matrix-of-U}) by $\supp^k(x)$. Notice, the number of
rows of the matrix depends on $k$. Let $Y\subset X_{[N]}$ be any subset
of cardinality $l$ and let $z\in X_{[N]}\setminus Y$. Assuming $l\geq 2$
we prove by contradiction that
\begin{equation}\label{eq:induction}
\begin{array}{c}
\supp^k(z)\subset\bigcup_{y\in Y}\supp^k(y)\\
\implies\forall x\in Y:
\supp^{k-1}(z)\subset\bigcup_{y\in Y\setminus\{x\}}\supp^{k-1}(y).
\end{array}
\end{equation}
The conclusion of (\ref{eq:induction}) says that for all $x\in Y$ and all
subsets $A\subset[N]$ of cardinality $k-1$ there exists
$y\in Y\setminus\{x\}$ such that $z_A=y_A$. The negation asserts the
existence of $x\in Y$
and $A\subset[N]$ of size $k-1$ such that for all $y\in Y\setminus\{x\}$
we have $z_A\neq y_A$. Hence, for all subsets $B\subset[N]$,
$B\supset A$ of size $k$ and for all $y\in Y\setminus\{x\}$ we have
$z_B\neq y_B$. The premise of (\ref{eq:induction}) then shows
$z_B=x_B$. Since one point of $B$, the one not in $A$, is free to move
within $[N]$, we get $z=x$ and the contradiction $z\in Y$ follows.
\par
Again by contradiction we prove the lemma. If a subset $Y\subset X_{[N]}$
of cardinality $l$ is not $k$-feasible
then there exists $z\in X_{[N]}\setminus Y$ such that the premise of
(\ref{eq:induction}) is true. Applying (\ref{eq:induction}) $l-1$ times
shows for all $x\in Y$ that $\supp^{k-l+1}(z)=\supp^{k-l+1}(x)$ holds.
Since $k-l+1\geq 1$ holds, this proves $z=x$ and contradicts $z\not\in Y$.
\hspace*{\fill}$\square$\\
%
%
%
%
\section{Divergence from a Gibbs family}
\label{sec:fundamentals}
\par
We prove that the correlation $c_k$ is the divergence from the Gibbs family
$\cE_k$ of $k$-local Hamiltonians. Thereby we use the fact that the divergence
from a Gibbs family is simply a
difference of von Neumann entropies, which in the case of the Gibbs family
$\cE_k$ already equals $c_k$ by definition. 
\par
This result is based on our work on information convergence 
\cite{Weis-Knauf,Weis-topology}. An almost identical result in terms of the 
irreducible correlation was proved in \cite{Weis-MaxEnt14}. Information
convergence has been studied in infinite-dimensional settings, too
\cite{Csiszar,Harremoes,Shirokov}.
\par
We consider a C*-algebra $\cA\subset\cM_d$, $d\in\bN$, containing
the identity $\id_d$. The {\it state space} of $\cA$ is the set of
all states in $\cA$ and will be denoted by $\cS_\cA$.
Let $\cH\subset\cA\her$ be a (real) subspace of self-adjoint matrices.
Using the map $\cA\her\to\cS_\cA$, $R(a)=e^a/{\rm tr}(e^a)$, we
define a {\it Gibbs family} $\cE:=R(\cH)$. In statistical physics,
the elements of $\cH$ are called {\em Hamiltonians\/} or {\em energies\/}.
\par
The {\it rI-closure} of a subset $X\subset\cS_\cA$ is defined by
\[
\cl(X):=\{\rho\in\cS_\cA\mid\inf_{\sigma\in X}D(\rho\|\sigma)=0\}.
\]
The acronym {\it rI} stands for {\it reverse information} where
{\it reverse} refers to the argument order of the divergence
\cite{Csiszar-Matus}.
The rI-closures of Gibbs families are studied in \cite{Weis-topology} 
where it is shown that for every state $\rho\in\cS_\cA$ exists a unique 
state in $\cl(\cE)$, denoted $\pi_\cE(\rho)$, such that
$\langle h,\rho\rangle=\langle h,\pi_\cE(\rho)\rangle$ holds for
all $h\in \cH$, see Sec.~3.3 and Coro.~3.9 in \cite{Weis-topology}.
The {\it Pythagorean theorem}, see Sec.~3.4 and Coro.~3.9 in 
\cite{Weis-topology}, says that for every $\rho\in\cS_\cA$ and
for every $\sigma\in\cl(\cE)$
\begin{equation}\label{eq:Pythagorean-theorem}
D(\rho\|\sigma)=D(\rho\|\pi_\cE(\rho))+D(\pi_\cE(\rho)\|\sigma)
\end{equation}
holds. Let us denote the divergence from $\cE$ by
\begin{equation}\label{eq:entropy-dist}
\dd_\cE(\rho):=\inf\{D(\rho\|\sigma)\mid\sigma\in\cE\},
\quad\rho\in\cS_\cA.
\end{equation}
The {\it projection theorem}, see Sec.~3.5 in \cite{Weis-topology},
says that for every $\rho\in\cS_\cA$ we have
\begin{equation}\label{eq:projection-theorem}
\dd_\cE(\rho)=D(\rho\|\pi_\cE(\rho))
=\min\{D(\rho\|\sigma)\mid\sigma\in\cl(\cE)\}
\end{equation}
and $\pi_\cE(\rho)$ is the unique local minimizer of the divergence
$D(\rho\|\,\cdot\,)$ on $\cl(\cE)$.
The theorems (\ref{eq:projection-theorem}) and
(\ref{eq:Pythagorean-theorem}) are topological extensions of results in
information geometry, see for example \cite{Petz,Amari-Nagaoka}, and
non-commutative extensions of results in probability theory, see for
example \cite{Csiszar-Matus}. The rI-closure in $\cS_\cA$ is in fact a
topological closure \cite{Weis-topology} but this is not essential now.
We come back to continuity issues later.
\par
For our purposes of maximum entropy states it suffices to draw two
consequences from the above statements. The first consequence, also
observed in Sec.~3.4 in \cite{Weis-topology}, follows from eq.\
(\ref{eq:Pythagorean-theorem}) by taking $\sigma=\id_d/\tr(\id_d)$
and using $D(\rho\|\id_d/\tr(\id_d))=\log(d)-H(\rho)$. The distance-like
properties of $D$ proves for all $\rho\in\cS_\cA$ that
\begin{equation}\label{eq:maxent}
\pi_\cE(\rho)=
{\rm argmax}\{H(\tau)\mid\tau\in\cS_\cA, \forall h\in \cH :
\langle h,\tau\rangle=\langle h,\rho\rangle\}.
\end{equation}
So $\pi_\cE$ is the maximum-entropy state under the constraints in
(\ref{eq:maxent}).
Secondly, the Pythagorean theorem proves, using the equality
$\dd_\cE(\rho)=D(\rho\|\pi_\cE(\rho))$ in (\ref{eq:projection-theorem})
that
\begin{equation}\label{eq:de=diffH}
\dd_\cE(\rho)=H(\pi_\cE(\rho))-H(\rho).
\end{equation}
The eq.\ (\ref{eq:de=diffH}) was also observed in \cite{Weis-MaxEnt14},
eq.\ (7).
\par
Let us now apply these results to the composite quantum system in
Sec.~\ref{sec:intro} where the algebra is
$\cA_{[N]}=\cA_1\otimes\cdots\otimes\cA_N$.
\begin{Cor}\label{cor:c=d}
For all $k=1,\ldots,N$ we have $c_k=\dd_{\cE_k}$.
\end{Cor}
{\it Proof:}
In view of (\ref{eq:maxent}) and (\ref{eq:de=diffH}) it suffices to
show for any state $\rho$ in $\cA_{[N]}$ that the constraint set
in (\ref{eq:maxent}) equals the set of states $\sigma$ in $\cA_{[N]}$
which have on all $k$-party subsystems the same marginals as $\rho$.
This is an easy calculation.
\hspace*{\fill}$\square$\\
\par
Needless to say that Coro.~\ref{cor:c=d} extends to more general
interaction patterns as provided by the notion of hierarchical
model in the next section. The divergence from a hierarchical model
has therefore, by applying the maximum-entropy principle like
in Sec.~\ref{sec:maxent-to-div-intro}, an interpretation
as correlation quantity.
\par
The above discussion allows to have a geometric view of the decomposition
by particle numbers
\[
c_1=C_2+\cdots+C_N
\]
of the total correlation $c_1$ in term of irreducible correlation
$C_k$. The irreducible correlation can be written in the form
($2\leq k\leq N$)
\[
C_k(\rho)=c_{k-1}(\rho)-c_k(\rho)
=D(\rho\|\pi_{\cE_{k-1}}(\rho))-D(\rho\|\pi_{\cE_k}(\rho))
=D(\pi_{\cE_k}(\rho)\|\pi_{\cE_{k-1}}(\rho))
\]
for all states $\rho$ in $\cA_{[N]}$ because of
(\ref{eq:Pythagorean-theorem}). Notice that $\cH_{k-1}\subset\cH_k$ holds
for the spaces of local Hamiltonians $\cH_{k-1},\cH_k$.
An analogous decomposition exists for any sequence
$H_1\subset H_2\subset\cdots\subset H_k\subset\cM_d$, $d\in\bN$,
of subspaces of hermitian matrices.
\par
Let us emphasize that the divergence from a Gibbs family $\cE$ is not
always continuous. This happens when the rI-closure $\cl(\cE)$ is not
norm closed \cite{Weis-topology}. The simplest example where the
divergence is discontinuous is a two-dimensional Gibbs family in the
algebra $\cM_3$ of $3\times 3$ matrices which is discussed in
\cite{Weis-Knauf,Weis-topology}. Discontinuities exists also
in the many-party correlation measures $c_k$. The total correlation
$c_1$ is continuous since it is of the form (\ref{eq:c1=H}) and because
the von Neumann entropy is continuous \cite{Wehrl1978}. The $2$-party
correlation $c_2$ of three qubits
is discontinuous at the GHZ state (and zero for almost all pure
states), see the discussions in \cite{Weis-MaxEnt14,RSSW}.
%
%
%
\section{Hierarchical models of quantum states}
\label{sec:hierarchical-models}
\par
Here we generalize the Gibbs families $\cE_k$ of $k$-local Hamiltonians 
from $k$-party interactions
to more complex interaction structures between subsystems. Similar
concepts appear in
theoretical biology and other disciplines, and have been abstractly
studied under the name of {\it hierarchical model}, see \cite{Lauritzen},
Chap.~4.3 and App.~B.2. We compute the dimension of a hierarchical
model. We also discuss a basis of the matrix algebra $\cM_d$.
\par
We consider the composite system from Sec.~\ref{sec:intro} with algebra
$\cA_{[N]}=\cA_1\otimes\cdots\otimes\cA_N$. Recall that
$\cA_k\subset\cM_{n_k}$ contains the identity matrix $\id_{n_k}$
of the size $n_k$, $k\in[N]$. To a non-empty subset
$v\subset[N]$ we associate the {\it factor space}
$\cF_v:=\cA_v\otimes\id_{[N]\setminus v}$ by embedding the algebra
$\cA_v=\bigotimes_{k\in v}\cA_{k}$ into $\cA_{[N]}$. We set
$\cF_\emptyset := {\rm span}_\bC(\id_{[N]})$. So
$\dim_\bC(\cF_v)= \prod_{k\in v} \dim_\bC(\cA_k)$, and
$\cF_w\subset\cF_v$ for $w\subset v$.
\par
The {\it pure factor space}
$\tilde\cF_v\subset \cF_v$ is then defined to be the maximal subspace
orthogonal (w.r.t.\ Hilbert-Schmidt inner product) to all $\cF_w$ with
$w\subsetneq v$. So
$\cF_v =\bigoplus_{w\subset v}\tilde\cF_w$,
and by M\"obius inversion applied to the dimensions of the subspaces, see for example
App.~A.3 in \cite{Lauritzen},
\begin{equation}\label{min:one}\textstyle
\dim_\bC(\tilde\cF_v) = \prod_{k \in v}\big(\dim_\bC(\cA_k)-1\big).
\end{equation}
\par
A basis of $\cA_{[N]}$ compatible with the
decomposition $\cA_{[N]}=\bigoplus_{v\subset[N]}\tilde\cF_v$ 
can be constructed from any family of orthonormal bases $B^{(k)}$
of $\cA_k$, such that $\frac{\id_{\{k\}}}{\sqrt{n_k}}\in B^{(k)}$,
$k\in[N]$. Then
\[\textstyle
\left\{ \bigotimes_{k=1}^N b_k \mid b_m\in B^{(m)}, m\in[N]\right\}
\]
is an orthonormal basis of $\cA_{[N]}$ and for $v\subset[N]$ we have
\[
\tilde\cF_v =
{\rm span} \left\{ \bigotimes_{k = 1}^N b_k \mid
b_m=\frac{\id_{\{m\}}}{\sqrt{n_m}} \text{ iff }
m\not\in v, b_m\in B^{(m)}, m\in[N]\right\}.
\]
\par
Sometimes a concrete basis is needed. For a full matrix
algebra ${\mathcal M}_n$ we can use for $k,l=0,\dots,n-1$ the matrices
given (for $r,s = 1,\ldots,n$) by
\[\textstyle
\Big(E_{k,l}^{(n)}\Big)_{r,s} :=
\frac{1}{\sqrt{n}}
\Big( \exp\left( \pi i (r+s) \frac{k}{n}\right) \, \delta_{r - s + l}
+
\exp\big( \pi i (r+s -n ) \frac{k}{n}\big) \, \delta_{r - s + l -n}
\Big).
\]
\begin{Lem}\label{lem:andreas-basis}
$\left\{ E_{k,l}^{(n)} \mid k,l\in\{0,\dots,n-1\}
\right\}\subset{\mathcal M}_n$ is an orthonormal basis of
${\mathcal M}_n$. The adjoints are
$E_{k,0}^{(n)}{}^*=E_{n-k,0}^{(n)}$,
$E_{0,l}^{(n)}{}^*=E_{0,n-l}^{(n)}$ and
$E_{k,l}^{(n)}{}^*=(-1)^{n+k+l}E_{n-k,n-l}^{(n)}$ for
$k,l=1,\ldots,n-1$.
\end{Lem}
{\it Proof:}
For $k, l, k', l' \in \{0,\dots,n-1\}$
\begin{eqnarray*}
\left\langle E_{k,l}^{(n)} , E_{k',l'}^{(n)} \right\rangle
   & = & \sum_{r,s = 1}^n \left( E_{k,l}^{(n)} \right)_{r,s} \overline{\left( E_{k', l'}^{(n)} \right)_{r,s}}  \\
   & = & \frac{1}{n} \sum_{r,s = 1}^n
   \Big[ \exp\big(\pi \, i \, (r + s)(k - k')/n \big) \, \delta_{r - s + l} \, \delta_{r - s + l'} \;\; + \;\;  \\
   &    & \qquad \exp\big( \pi \, i \, (r + s -n)(k - k')/n \big) \,
            \delta_{r - s + l -n} \, \delta_{r - s + l' - n} \Big] \\
   & = & \frac{1}{n} \, \delta_{l, l'}\; \sum_{r = 1}^n
   \exp\big( \pi \, i \, (2 r + l)(k - k')/n\big)
   = \delta_{l,l'} \, \delta_{k,k'}.
\end{eqnarray*}
As the set has size $n^2$, this shows the claim. The following
adjoints appear. One has
$E_{0,0}^{(n)}=\frac{1}{\sqrt{n}}\id_n$. For $k=1,\ldots,n-1$ and
coefficients $r,s=1,\ldots,n$
\begin{eqnarray*}
\left(E_{k,0}^{(n)}\right)^*_{r,s}
& = & \frac{1}{\sqrt{n}}
\ov{\exp\left(\pi i (r+s)k/n\right)}\delta_{r-s}
 = \frac{1}{\sqrt{n}}\exp\left(
-\pi i (r+s)k/n\right)\delta_{r-s} \\
& = & \frac{1}{\sqrt{n}}\exp\left(
\pi i (r+s)(n-k)/n\right)\delta_{r-s}
 = \left(E_{n-k,0}^{(n)}\right)_{r,s}
\end{eqnarray*}
holds and for $l=1,\ldots,n-1$ it is immediate that
$E_{0,l}^{(n)}{}^*=E_{0,n-l}^{(n)}$. For $k,l=1,\ldots,n-1$ and
coefficients $r,s=1,\ldots,n$  one has
\begin{eqnarray*}
\left(E_{k,l}^{(n)}\right)^*_{r,s}
& = & \frac{1}{\sqrt{n}}\Big(
\exp\left(-\pi i(r+s-n)k/n\right)\delta_{r-s+n-l}
\;\; + \;\;\\
& & \qquad
\exp\left(-\pi i(r+s)k/n\right)\delta_{r-s-l}\Big)\\
& = & \frac{1}{\sqrt{n}}\Big(
(-1)^{k+r+s}\exp\left(\pi i(r+s)(n-k)/n\right)\delta_{r-s+n-l}
\;\; + \;\; \\
& & \qquad \hspace{1ex}(-1)^{n+k+r+s}\exp\left(\pi
i(r+s-n)(n-k)/n\right)\delta_{r-s-l}\Big)\\
& = & (-1)^{n+k+l}\left(E_{n-k,n-l}^{(n)}\right)_{r,s}.
\end{eqnarray*}
\hspace*{\fill}$\square$\\
\par
One way to compute a self-adjoint basis out of the basis
$\{E_{k,l}^{(n)}\}_{k,l=0}^{n-1}$ of $\cM_n$, $n\in\bN$, in
Lemma~\ref{lem:andreas-basis}, is to use their symmetry under hermitian
conjugation. Orbits have length one or two. Thus the transformation of
basis matrices $E$ to pairs of matrices $E+E^*$ and ${\it i}(E-E^*)$
produces exactly $n^2$ pairwise orthogonal non-zero self-adjoint matrices.
This symmetrization is different compared to the basis (3.2) in
\cite{Petz}, where only real hermitian matrices appear which are
either diagonal or which have only two non-zero entries. In contrast\\
\centerline{%
$E_{0,1}^{(3)}+(E_{0,1}^{(3)})^*=\frac{1}{\sqrt{3}}
\scalebox{0.71}{$\left(\begin{array}{ccc}
0 & 1 & 1\\
1 & 0 & 1\\
1 & 1 & 0
\end{array}\right)$}$.}\\
\par
Returning to the subject of hierarchical models, let $U\subset 2^{[N]}$ be a
class of subsets of $[N]$.
Differing from common terminology, we will call $U$ a {\em hypergraph\/}
on $[N]$ if
\[
       v \in U, \; w \subset v \;\; \Rightarrow \;\; w \in U, \qquad
       \mbox{and} \qquad \bigcup_{v\in U} v = [N].
\]
We consider a hypergraph $U$ on $[N]$ and define the
{\it hierarchical model subspace}
$\tilde\cF_U := \bigoplus_{v \in U} \tilde\cF_v$. The
{\it hierarchical model} $\cE_U$ of $U$ is defined as the Gibbs family
\begin{equation}\label{eq:qm-hierarchical}
\cE_U:=R(\tilde\cF_U \cap \cA_{[N]}\her).
\end{equation}
Of particular interest are the hypergraphs
$U_k = \bigcup_{\ell= 0}^k{\binom{[N]}{\ell}}$ where ${\binom{[N]}{\ell}}$
denotes the class of subsets of $[N]$ having $\ell$ elements. The Gibbs
family $\cE_k$ of the $k$-local Hamiltonians (\ref{eq:Ek})
is the hierarchical model of the hypergraph $U_k$. For example, the 
{\it independence model} $\cE_1$ is the hierarchical model of the hypergraph 
$\{\emptyset, \{1\},\ldots,\{N\}\}$.
\par
We now compute dimensions. The relative interior of
a subset of $\cA\her$ is the interior of the subset in its affine hull.
\begin{Pro}\label{prop:dim}
Let $U$ be a hypergraph on $[N]$. Then the hierarchical model subspace
$\tilde\cF_U$ has dimension
\[\textstyle
{\rm dim}_{\bC} ( \tilde\cF_U )
= \sum_{v \in U} \prod_{i \in v} \big(\dim_{\bC}(\cA_i)-1\big).
\]
The subspace of hermitian matrices satisfies
 ${\rm dim}_{\bR} ( \tilde\cF_U \cap \cA_{[N]}\her )
={\rm dim}_{\bC} ( \tilde\cF_U )$ and the Gibbs family
${\mathcal E}_U$ has dimension
${\rm dim}_{\bR} ( {\mathcal E}_U )={\rm dim}_{\bC} ( \tilde\cF_U )-1$.
\end{Pro}
{\it Proof:}
By the definition of hypergraphs and by (\ref{min:one}) we have for all
$v\subset[N]$
\[
{\rm dim}_\bC \big( \tilde\cF_v \big)
=
\prod_{i \in v} \big(\dim_{\bC}(\cA_i)-1\big).
\]
A complex *-invariant subspace of $\cA$ is a direct sum of two copies of
the real subspace of its self-adjoint elements. Therefore
\[
{\rm dim}_\bR(\tilde\cF_v\cap \cA\her)
= {\rm dim}_\bC(\tilde\cF_v).
\]
By definition, the hypergraph $U$ contains $\emptyset$ and
$\tilde\cF_U=\tilde\cF_\emptyset\oplus V$ is the direct sum of
$\tilde\cF_\emptyset = {\rm span}_\bC (\id_{[N]})$ and of its orthogonal
complement, denoted $V$. Clearly $R(\tilde\cF_U \cap\cA\her)=R(V\cap\cA\her)$
holds. If $W\subset\cA\her$ is a codimension one subspace not containing
the identity $\id_\cA$, then $R|_W$ is a diffeomorphism to the relative
interior of $\cS_{\cA_{[N]}}$, see Prop.~6.1.2 in \cite{Weis-topology}.
Hence $\dim_\bR(\cE_U)=\dim_\bR(V)$ completes the proof.
\hspace*{\fill}$\square$\\
%
%
%
\section{The multi-information}
\label{sec:multi-information}
\par
Here we consider the total correlation $c_1$ and relations between
the independence model $\cE_1$ and the set of product states $\cF_1$
defined in (\ref{eq:F1}). Among others, we prove for every state
$\rho$ in $\cA_{[N]}$ that $c_1(\rho)$ is the multi-information
\begin{equation}\label{eq:multi-info}\textstyle
I(\rho):=\sum_{i\in[N]}H(\rho_{\{i\}})-H(\rho).
\end{equation}
This statement follows from Coro.~\ref{cor:c=d} and Thm.~\ref{thm:d=I}
and was claimed in (\ref{eq:c1=H}).
\begin{Thm}\label{thm:d=I}
We have $\cF_1=\cl(\cE_1)=\overline{\cE_1}$, that is the set of
product states is the rI-closure and the norm closure of the
independence model. We have $\dd_{\cE_1}=I$, that is the divergence
from the independence model is the multi-information.
\end{Thm}
{\it Proof:}
We prove $\cF_1\subset\cl(\cE_1)$. Let 
$\rho=\rho_{\{1\}}\otimes\cdots\otimes\rho_{\{N\}}$ be a product state
in $\cA_{[N]}$. 
It is shown in Thm.~5.18.5 in \cite{Weis-topology} that each individual 
factor $\rho_{\{i\}}$ lies in the rI-closure of the relative interior of the state
space $\cS_{\cA_i}$, which is the set of all invertible density matrices
in $\cS_{\cA_i}$. So there exist sequences
$(\rho_i^{(n)})_{n\in\bN}\subset\cS_{\cA_i}$ of invertible states such
that $\lim_{n\to\infty}D(\rho_{\{i\}}\|\rho_i^{(n)})=0$, $i\in[N]$. It
follows
\[
D\big(\rho\|\rho_1^{(n)}\otimes\cdots\otimes\rho_N^{(n)}\big)
=D\big(\rho_{\{1\}}\|\rho_1^{(n)}\big)+\cdots+D\big(\rho_{\{N\}}\|\rho_N^{(n)}\big)
\ \stackrel{n\to\infty}{\longrightarrow}\ 0.
\]
Since $\rho_1^{(n)}\otimes\cdots\otimes\rho_N^{(n)}\in\cE_1$ for all
$n\in\bN$ this proves $\rho\in\cl(\cE_1)$. The inclusion
$\cl(\cE_1)\subset\ov{\cE_1}$ follows from the Pinsker inequality
\cite{Petz2008}. The inclusion
$\ov{\cE_1}\subset\cF_1$ follows because $\cE_1\subset\cF_1$ and
because $\cF_1$ is norm closed since it is the image of the cartesian 
product of compact state spaces $\cS_{\cA_i}$, $i\in[N]$, under the
continuous tensor product map 
$(\rho_1,\ldots,\rho_N)\mapsto\rho_1\otimes\cdots\otimes\rho_N$.
This completes the proof of $\cF_1=\cl(\cE_1)=\overline{\cE_1}$.
\par
Now let $\rho$ be an arbitrary state in $\cA_{[N]}$, not necessarily
equal to the product of its marginals 
$\sigma:=\rho_{\{1\}}\otimes\cdots\otimes\rho_{\{N\}}$.
A short computation proves that $\sigma$ is the unique global
minimizer of the divergence $D(\rho\|\,\cdot\,)$ on $\cF_1$, see
\cite{Modi}, Lemma~1. Since $\cF_1=\cl(\cE_1)$ holds, the
projection theorem (\ref{eq:projection-theorem}) proves first that
$\sigma$ is the state $\pi_{\cE_1}(\rho)$ defined in 
Sec.~\ref{sec:fundamentals} and second that 
$\dd_{\cE_1}(\rho)=D(\rho\|\sigma)$ holds. The identity 
$D(\rho\|\sigma)=I(\rho)$ is very easy to compute and completes the proof. 
\hspace*{\fill}$\square$\\
%
%
%
\section{Local maximizers of the divergence}
\label{sec:local-maximizers}
\par
We evaluate a support bound for a local maximizer of the divergence 
from a Gibbs family and we recall a second condition for a local maximizer. 
The conditions go back to the work of one of us \cite{Ay} 
in probability theory and have been extended to quantum states in  
\cite{Weis-Knauf,Weis-topology}.
\par
The support bound is derived from a bound on the face dimensions of the 
state space $Z:=\cS_{\cA_{[N]}}$ which is a compact and convex set. We 
sketch the proofs in \cite{Ay,Weis-topology}.
A {\em face} of $Z$ is any convex subset $F\subset Z$ such that every 
segment in $Z$ which meets $F$ with an interior point lies in $F$. 
A face which is a singleton is called {\em extremal point}.
For every state $\rho$ in $Z$ exists a unique face $F(\rho)$ of $Z$ 
such that $\rho$ lies in the relative interior of $F(\rho)$. If an affine
space $A$ contains $\rho$ then $\rho$ lies in the relative interior
of the intersection $A\cap F(\rho)$. See for example \cite{Rockafellar}
for these statements.
\par
We consider a C*-algebra $\cA\subset\cM_d$, $d\in\bN$, with $\id_d\in\cA$.
Like in Sec.~\ref{sec:fundamentals} we define a Gibbs family $\cE=R(\cH)$
in terms of a space $\cH\subset\cA\her$ of self-adjoint matrices. For any 
state $\rho$ in $\cA$ we consider the affine space
\[
A:=\{a\in\cA\her\mid\forall h\in \cH :
\langle h,a\rangle=\langle h,\rho\rangle\}
\]
and the convex set $A\cap F(\rho)$ which contains $\rho$ in its
relative interior. The divergence from $\cE$ is by (\ref{eq:de=diffH})
of the form 
\[
\dd_\cE(\rho)=H(\pi_\cE(\rho))-H(\rho).
\]
The first term is constant on $A\cap Z$ and the von Neumann entropy $H$ is
strictly concave on $Z$, see for example \cite{Wehrl1978}, so $\dd_\cE$
is strictly convex on $A\cap Z$. If $\rho$ is a local maximizer
of $\dd_\cE$ on $Z$ then $\rho$ is a local maximizer on the relative
interior $X$ of $A\cap F(\rho)$. By the strict convexity of $\dd_\cE$ the
local maximizer $\rho$ must be an extremal point of $X$. Since $X$ is 
relative open this proves, see \cite{Ay}, Prop.~3.2, that $A\cap F(\rho)$ 
is a singleton. Now 
\begin{equation} \label{dimineq}
\dim_\bR(F(\rho))\leq\dim_\bR(\cE)
\end{equation}
follows, see \cite{Weis-topology}, Prop.~6.17.
\par
The inequality (\ref{dimineq}) can be expressed in terms of the rank of a local
maximizer. Two extreme cases are discussed in Rem.~6.18 in
\cite{Weis-topology}: The classical algebra of diagonal matrices
$\cA\cong\bC^d$, where (\ref{dimineq}) becomes
\begin{equation}\label{eq:rk-E-classical}
{\rm rk}(\rho)\leq\dim_\bR(\cE)+1
\end{equation}
and the full matrix algebra
$\cA=\cM_d$, where  (\ref{dimineq}) becomes
\begin{equation}\label{eq:rk-E-quantum}
{\rm rk}(\rho)\leq\sqrt{\dim_\bR(\cE)+1}.
\end{equation}
\par
Let us evaluate these bounds for a hierarchical model $\cE_U$
based on a
hypergraph $U$ on $[N]$. Prop.~\ref{prop:dim} then shows
\[\textstyle
{\rm dim}_\bR \left( {\mathcal E}_U \right)
= \sum_{v \in U \atop v \not= \emptyset}
\prod_{k \in v} \big(\dim_\bC(\cA_k)-1\big).
\]
In the classical case of diagonal matrices
$\cA_{[N]}\cong\bC^{n_1}\otimes\cdots\otimes\bC^{n_N}$
the state space $\cA_{[N]}$ is a probability simplex.
A probability distribution $p$ which is a local maximizer of the
divergence from $\cE_U$ satisfies by (\ref{eq:rk-E-classical}) the bound
\begin{equation}\label{clbound}\textstyle
   |{\rm supp}(p)| \leq
   \sum_{v \in U \atop v \not= \emptyset}
   \prod_{i \in v}( {n_i} -1 ) + 1\,.
\end{equation}
In the quantum case
$\cA_{[N]}=\cM_{n_1}\otimes\cdots\otimes\cM_{n_N}$ a local maximizer
$\rho$ of the divergence from $\cE_U$ satisfies by (\ref{eq:rk-E-quantum})
bound
\begin{equation}\label{qmbound}\textstyle
{\rm rk}(\rho)
\leq \sqrt{
\sum_{v \in U \atop v \not= \emptyset}
\prod_{i \in v} ( {n_i}^2 -1 ) + 1}\,.
\end{equation}
\par
It is very interesting to derive the corresponding bounds for the many-party 
correlation $c_k$ given uniform unit sizes $n\in\bN$. Recall
from Coro.~\ref{cor:c=d} that $c_k$ is the divergence from the Gibbs
family $\cE_k$ of the $k$-local Hamiltonians whose hypergraph $U_k$ is 
defined in the paragraph of (\ref{eq:qm-hierarchical}). A local maximizer $p$
(classical case) resp.\ $\rho$ (full matrix algebra) of $c_k$ satisfies
by (\ref{clbound}) resp.\ (\ref{qmbound}) the bound
\begin{equation}\textstyle\label{eq:equal-units-cl}
|{\rm supp}(p)|\leq\sum_{i=1}^k\binom{N}{i}(n-1)^i+1
\quad\mbox{resp.}\quad
{\rm rk}(\rho)\leq\sqrt{\sum_{i=1}^k\binom{N}{i}(n^2-1)^i+1}.
\end{equation}
The bounds for the multi-information $I=c_1$ are
$N(n-1)+1$ resp.\ $\sqrt{N(n^2-1)+1}$.
\par
For curiosity we mention a second characterization of a local maximizer
$\rho$ of the divergence from a Gibbs family $\cE=R(\cH)$, defined as
above. Namely, $\rho$ must have a special form. A projection in $\cA$
is a matrix such that $p=p^2=p^*$ holds. One of us has shown in
\cite{Weis-topology}, Secs.~3.3 and~3.5, 
that the state $\pi_\cE(\rho)\in\cl(\cE)$
defined in Sec.~\ref{sec:fundamentals} is of the form
$q e^{qa_\rho q}/\tr( qe^{qa_\rho q})$ for some self-adjoint matrix
$a_\rho\in\cH$ and projection $q$. Surprisingly, the Coro.~6.19 in 
\cite{Weis-topology} shows that
a local maximizer $\rho$ of the divergence from $\cE$ is itself of
the form $\rho=p e^{pa_\rho p}/\tr(pe^{pa_\rho p})$ for a projection
$p\in\cA$. We have proved the case $q=\id_d$ already in 
\cite{Weis-Knauf} by computing partial derivatives in a straight 
forward generalization of the classical case \cite{Ay}. 
Further results in this direction have been found in  \cite{Matus2007}.
%
%
%
%
\section{Separable qubit states and maximizers of the mutual information}
\label{sec:global-maximizers}
\par
We have studied global maximizers of the multi-information of probability
distributions in \cite{Ay-Knauf}. For example, a classification was proved for
global maximizers. If the units are ordered
by their size, such that $n_1\leq\cdots\leq n_N$, then the bound
of the multi-information (\ref{eq:multi-info}) is
\[\textstyle
I(p)\leq\sum_{i=1}^{N-1}\log(n_i),
\quad p\in\cS_{\cA_{[N]}}\cong\Delta(n_1\times\cdots\times n_N)
\]
for probability distributions $p$. For example, two classical bits have
$\log(2)=1$ bit of maximal mutual information. The example of two maximally
entangled qubits, for example the Bell state
$\tfrac{1}{\sqrt{2}}(|00\rangle+|11\rangle)$, shows that quantum systems can
break the classical bound. This is a reason why some of the basic ideas in
\cite{Ay-Knauf} do not apply to the quantum setting of full matrix algebras,
$\cA_i=\cM_{n_i}$, $i\in[N]$.
\par
Here we show that some arguments from \cite{Ay-Knauf} are helpful in the
maximization of multi-information on the separable states. By definition, a
state in $\cA$ is {\it separable} if it is a convex combination of product
states $\rho_1\otimes\cdots\otimes\rho_N$. A state which is not separable is
{\it entangled} \cite{Nielsen,Bengtsson-Zyczkowski}. We restrict the
discussion to the simplest case of a bipartite system ($N=2$) of two qubits
$\cA_1=\cA_2=\cM_2$ where the multi-information (\ref{eq:multi-info}) is
known as {\it mutual information}
\begin{eqnarray}
I(\rho)=H(\rho_{\{1\}})+H(\rho_{\{2\}})-H(\rho),
\quad \rho\in\cS_\cA,
\quad \cA=\cA_1\otimes\cA_2.
\end{eqnarray}
A state is {\it classically correlated} \cite{Modi} if it can be
diagonalized by local unitaries that is, matrices in the subgroup
$U(2)\times U(2)\subset U(4)$. This class of states has been
discussed earlier in the literature in the context of quantum discord
\cite{OllivierZurek2001}.
\begin{Thm}\label{th 1}
For arbitrary separable two-qubit state $\rho$, its mutual information is bounded
by  $I(\rho)\leq \log(2)$. The equality holds if and only if $\rho$ is
local unitary equivalent to $\frac{1}{2}(|0\rangle\langle0|\otimes|0\rangle\langle0|
+|1\rangle\langle 1|\otimes|1\rangle\langle 1|)$. In particular, all separable
maximizers of the mutual information of two qubits are classically correlated.
\end{Thm}
{\it Proof:}
If $\rho$ is separable, then $H(\rho_{\{i\}})\leq H(\rho)$, $i=1,2$, holds, see
\cite{M. A. Nielsen}. So we have
\begin{eqnarray}
I(\rho)\leq \min\{H(\rho_{\{1\}}),\ H(\rho_{\{2\}})\}.
\end{eqnarray}
For qubit states $\rho_{\{1\}}$ and $\rho_{\{1\}}$, the maximum of the
von Neumann entropy is no more than $\log(2)$, which constrains the maximum of
mutual information $I(\rho)$ to $\log(2)$. So if
$I(\rho)$ reaches its maximum $\log(2)$, then $H(\rho_{\{i\}})$, $i=1,2$, also
reaches this maximum, which requires $\rho_{\{i\}}$ to be the maximally mixed state
$\frac{1}{2}\id_2$.
\par
Two-qubit mixed states with maximally mixed reduced states are
local unitary equivalent to {\it Bell-diagonal states}
\begin{eqnarray}\label{two-qubit Bell diag}
\rho=\sum_{i=1}^4 \lambda_i |\psi_i\rangle\langle\psi_i|,
\quad \lambda_1,\lambda_2,\lambda_3,\lambda_4\geq 0,
\quad \lambda_1+\lambda_2+\lambda_3+\lambda_4=1
\end{eqnarray}
with
$|\psi_1\rangle=\frac{1}{\sqrt{2}}(|00\rangle+|11\rangle)$, $|\psi_2\rangle=\frac{1}{\sqrt{2}}(|00\rangle-|11\rangle)$, $|\psi_3\rangle=\frac{1}{\sqrt{2}}(|01\rangle+|10\rangle)$, $|\psi_4\rangle=\frac{1}{\sqrt{2}}(|01\rangle-|10\rangle)$, see
\cite{O. Rudolph}. Note that $-H(\rho)$ is a strictly convex function of quantum
states, subsequently, the maximum of $I(\rho)$ on the convex set of separable
Bell-diagonal states is attained only on the extreme points of this convex set.
A Bell-diagonal state is separable if and only if $\lambda_i\leq \frac{1}{2}$
for $i=1,2,3,4$, see \cite{Horodecki,Lang-Caves}. We find
the extreme points of the set of separable Bell-diagonal states are
\begin{eqnarray}\label{two-qubit extreme points Bell diag}
\frac{1}{2}(|\psi_i\rangle\langle\psi_i| +|\psi_j\rangle\langle\psi_j|),
\quad i\neq j,
\quad i,j=1,2,3,4.
\end{eqnarray}
One can verify further that the mutual information of
all these extreme points is $\log(2)$. Therefore the separable two-qubit states
with maximum mutual information are all local unitary equivalent to the
quantum state in (\ref{two-qubit extreme points Bell diag}).
\par
Now we take a closer look at these maximizers. We find they are all classically
correlated, since
\begin{equation}\label{two-qubit extreme points classical}
\begin{array}{rcl}
\frac{1}{2}(|\psi_1\rangle\langle\psi_1|+|\psi_2\rangle\langle\psi_2|)&=&\frac{1}{2}(|0\rangle\langle0|\otimes|0\rangle\langle0|
+|1\rangle\langle 1|\otimes|1\rangle\langle 1|);\\
\frac{1}{2}(|\psi_1\rangle\langle\psi_1|+|\psi_3\rangle\langle\psi_3|)&=&\frac{1}{2}(|+\rangle\langle +|\otimes|+\rangle\langle +|+|-\rangle\langle -|\otimes|-\rangle\langle -|);\\
\frac{1}{2}(|\psi_1\rangle\langle\psi_1|+|\psi_4\rangle\langle\psi_4|)&=&\frac{1}{2}(|0^\prime\rangle\langle 0^\prime|\otimes|1^\prime\rangle\langle 1^\prime|+|1^\prime\rangle\langle 1^\prime|\otimes|0^\prime\rangle\langle 0^\prime|);\\
\frac{1}{2}(|\psi_2\rangle\langle\psi_2|+|\psi_3\rangle\langle\psi_3|)&=&\frac{1}{2}(|1^\prime\rangle\langle 1^\prime|\otimes|1^\prime\rangle\langle 1^\prime|+|0^\prime\rangle\langle 0^\prime|\otimes|0^\prime\rangle\langle 0^\prime|);\\
\frac{1}{2}(|\psi_2\rangle\langle\psi_2|+|\psi_4\rangle\langle\psi_4|)&=&\frac{1}{2}(|-\rangle\langle -|\otimes|+\rangle\langle +|+|+\rangle\langle +|\otimes|-\rangle\langle-|);\\
\frac{1}{2}(|\psi_3\rangle\langle\psi_3|+|\psi_4\rangle\langle\psi_4|)&=&\frac{1}{2}(|0\rangle\langle 0|\otimes|1\rangle\langle 1|+|1\rangle\langle 1|\otimes|0\rangle\langle0|),
\end{array}
\end{equation}
with
$|+\rangle=\frac{1}{\sqrt{2}}(|0\rangle+|1\rangle)$, $|-\rangle
=\frac{1}{\sqrt{2}}(|0\rangle-|1\rangle)$, $|0^\prime\rangle
=\frac{1}{\sqrt{2}}(|0\rangle+i|1\rangle)$, $|1^\prime\rangle
=\frac{1}{\sqrt{2}}(|0\rangle-i|1\rangle)$. Here $\{|+\rangle, |-\rangle \}$ and
$\{|0^\prime\rangle, |1^\prime\rangle\}$ are another two orthonormal bases of two
dimensional Hilbert space. From equations
(\ref{two-qubit extreme points classical}) it is direct to get that all the
maximizers are local unitary equivalent to
$\frac{1}{2}(|0\rangle\langle0|\otimes|0\rangle\langle0|
+|1\rangle\langle 1|\otimes|1\rangle\langle 1|)$.
\hspace*{\fill}$\square$\\
\par
We finish with a geometric discussion of Thm.~\ref{th 1}. Mutual information
is the relative entropy of a quantum state from its closest product state,
$I(\rho)=\min_{\pi\in\cF_1} D(\rho\|\pi)$, see \cite{Modi}.
Hence, the mutual information $I(\rho)$ can be regarded as the distance between
a quantum state and the set of product states $\cF_1$. In a two-qubit system,
the maximum distance between an arbitrary separable quantum state and the
set of product states $\cF_1$ is $\log(2)$. Thm.~\ref{th 1} shows the
farthest separable states from the set of product states $\cF_1$ are all local
unitary equivalent to $\frac{1}{2}(|0\rangle\langle0|\otimes|0\rangle\langle0|
+|1\rangle\langle 1|\otimes|1\rangle\langle 1|)$. These states are
classically correlated so they can not be used in the protocol of
entanglement distribution {\it via} separable states in \cite{Kay}.
\par
The Bell-diagonal states can be written as
$\rho=\frac{1}{4}(\id_4+\sum_{i=1}^3 t_i \sigma_i\otimes\sigma_i)$ with
$\sigma_i$ three Pauli operators. So a Bell-diagonal state is specified by
three real variables $t_1$, $t_2$, and $t_3$. One can show that a
Bell-diagonal state is separable if and only if $|t_1|+|t_2|+|t_3|\leq 1$
holds. Geometrically, the set of Bell-diagonal states is a tetrahedron and
the set of separable Bell-diagonal states is an octahedron, see
\cite{Horodecki,Lang-Caves} and Fig.~1 for a drawing. The four
vertices of the tetrahedron are Bell states $|\psi_i\rangle$ which are
maximally entangled, $i=1,2,3,4$. The six black vertices of the octahedron
are maximizers of the mutual information and they are classically correlated.
The center red point $\frac{1}{4}\id_4$ is the only product state in this
tetrahedron.
\begin{center}
\begin{figure}
\label{fig}
\resizebox{6cm}{!}{\includegraphics{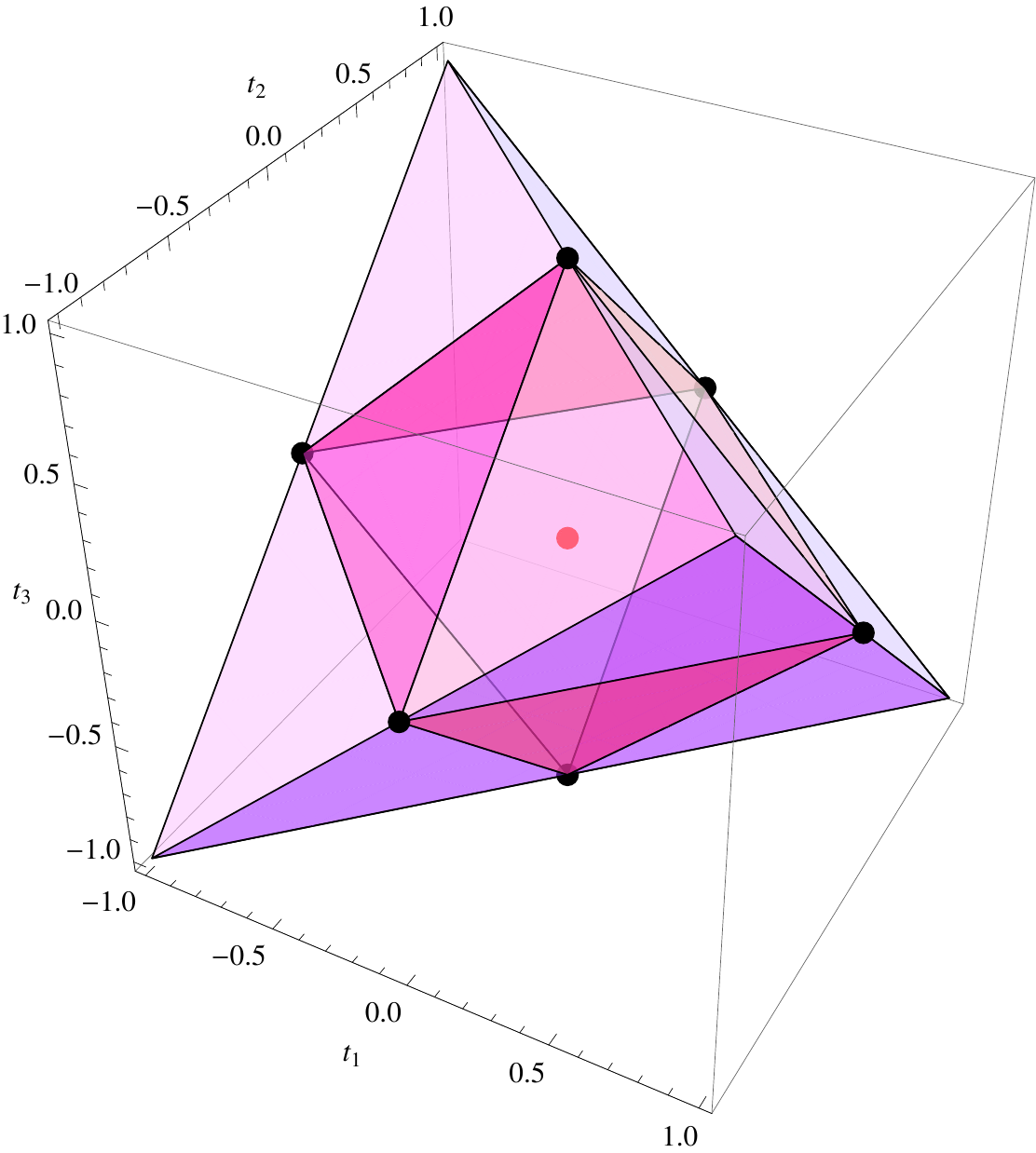}}
\caption{
Geometry of Bell-diagonal states. }
\end{figure}
\end{center}
%
%
%
%
%
\begin{Ack}
SW thanks Thomas Kahle for a helpful correspondence about factorization
of probability distributions. SW was partially supported by the DFG
projects ``Geometry and Complexity in Information Theory'' and
``Quantum Statistics:\ Decision problems and entropic functionals on
state spaces''. MJZ is supported by the NSF of China under Grant No.\
11401032 and SRF for ROCS, SEM.
\end{Ack}
%
%
%
%
%
%
%
\phantomsection
\addcontentsline{toc}{section}{References}
\bibliographystyle{nnapalike}

\newpage
\noindent
\begin{tabular}{l}
Stephan Weis\\
e-mail: maths@stephan-weis.info\\[1ex]
Max Planck Institute for \\
Mathematics in the Sciences \\
Inselstrasse 22 \\
D-04103 Leipzig \\
Germany 
\end{tabular}
\\[\baselineskip]
\begin{tabular}{l}
Andreas Knauf\\
e-mail: knauf@math.fau.de\\[1ex]
Department of Mathematics\\
Friedrich-Alexander-University\\
Erlangen-Nuremberg\\
Cauerstr.\ 11\\
D-91058 Erlangen\\
Germany
\end{tabular}
\\[\baselineskip]
\begin{tabular}{lll}
Nihat Ay\\
e-mail: nay@mis.mpg.de\\[1ex]
Max Planck Institute for & Department of Mathematics & Santa Fe Institute\\
Mathematics in the Sciences &  and Computer Science & 1399 Hyde Park Road\\
Inselstrasse 22 & Leipzig University & Santa Fe\\
D-04103 Leipzig & PF 10 09 20 & New Mexico 87501\\
Germany & D-04009 Leipzig & USA\\
 & Germany
\end{tabular}
\\[\baselineskip]
\begin{tabular}{l}
Ming-Jing Zhao\\
e-mail: zhaomingjingde@126.com\\[1ex]
Department of Mathematics\\
School of Science\\
Beijing Information Science and\\
Technology University\\
100192 Beijing\\
China
\end{tabular}
\end{document}